\documentclass[twoside]{article}
\usepackage{Proc_NTSE_18}

\begin{document}
\thispagestyle{plain}
\publref{CloudyBagBLFQ}

\begin{center}
{\Large \bf \strut
Dynamical Nucleon-Pion System via Basis Light-Front Quantization
\strut}\\
\vspace{10mm}
{\large \bf 
Weijie Du$^{a}$, Yang Li$^{a,b,c}$, Xingbo Zhao$^{d,e}$ and  James P. Vary{$^a$}}
\end{center}

\noindent{
\small $^a$\it Department of Physics and Astronomy, Iowa State University, Ames, Iowa 50010, USA } \\
{\small $^b$\it Hebei Key Laboratory of Compact Fusion, Langfang 065001, China} \\
{\small $^c$\it ENN Science and Technology Development Co., Ltd., Langfang 065001, China } \\
{\small $^d$\it Institute of Modern Physics, Chinese Academy of Sciences, Lanzhou 730000, China} \\
 {\small $^e$\it University of Chinese Academy of Sciences, Beijing 100049, China}

\markboth{
Weijie Du, Yang Li, Xingbo Zhao and James P. Vary}
{
Dynamical Nucleon-Pion System via Basis Light-Front Quantization} 

\begin{abstract}
We present the first application of the Basis Light-Front Quantization method to study a simple chiral model of the nucleon-pion system via an {\it ab initio}, non-perturbative, Hamiltonian approach. As a test problem, we consider the physical proton as the relativistic bound state of the nucleon-pion system. Based on the chiral model of the nucleon-pion system, we construct the mass-squared matrix of the system within our light-front basis representation. We obtain the proton's mass and the corresponding light-front wave function by solving the eigenvalue problem of the mass-squared matrix. With the resulting boost-invariant light-front wave function, we also compute the proton's parton distribution function.
\\[\baselineskip] 
\end{abstract}

\section{Introduction}
Developing a relativistic methodology that is broadly applicable to nuclear physics is important. Progress in this direction will be useful for studying high-momentum experiments of nuclear targets using exclusive, nearly exclusive or inclusive processes \cite{Hen:2016kwk,Aubert:1983xm,HCheng:1987}. One of the promising methods for such investigations is the Basis Light-Front Quantization (BLFQ) method \cite{Vary:2009gt}.

BLFQ is a non-perturbative, {\it ab initio} method, which treats relativistic quantum field theory via the Hamiltonian approach within the light-front (LF) formalism. BLFQ has been shown to be a promising tool in a range of applications, such as the electron anomalous magnetic moment \cite{Honkanen:2010rc,Zhao:2014xaa}, the positronium spectrum \cite{Wiecki:2014ola}, and the heavy quarkonium structure and radiative transitions \cite{Vary:2018pmv,Li:2015zda,Li:2017mlw,Tang:2018myz,Li:2018uif}. More recently, BLFQ has been applied successfully to the properties of the light mesons \cite{Jia:2018ary}, which are then extended to higher scales by QCD evolution \cite{Lan:2019vui}. This Hamiltonian approach has also been extended to develop a non-perturbative scattering framework through time-dependent BLFQ (tBLFQ) \cite{Zhao:2013cma,Zhao:2013jia,Chen:2017uuq}.

The LF quantization procedure for treating a chiral nucleon-pion ($N\pi$) model was first proposed by Miller \cite{Miller:1997cr,Miller:2000kv} in studying the $N\pi$ scattering and the nucleon-nucleon scattering via the perturbative approach. In this work, we present the first non-perturbative, {\it ab initio} treatment of the same chiral model. As a test problem, we consider the physical proton as the relativistic bound state of the $N\pi$ system. Using the BLFQ method, we compute the mass-squared matrix element of the $N\pi$ system within the LF basis representation. We then solve the eigenvalue problem of the resulting mass-squared matrix and obtain the proton's mass and LFWF. The proton's LFWF is boost-invariant and can be directly applied to compute the observables such as the parton distribution function (PDF). 

The outline of this paper is the following. We begin with the theory part in Sec. \ref{sec:SecTheory}, which introduces the elements of BLFQ, such as the derivation of the LF Hamiltonian density, our choice of the basis construction and truncation schemes, the derivation of the mass-squared matrix element in the basis representation, and the formalism of the observables in this work. We present the results of the proton's mass, LFWF and PDF in Sec. \ref{sec:ResultsDiscussions}. We conclude in Sec. \ref{sec:ConclusionOutlook}, where we also discuss our future plans.

\section{BLFQ approach to a chiral model}
\label{sec:SecTheory}
\subsection{Hamiltonian dynamics}
The dynamical $N\pi$ system can be evaluated from the eigenvalue equation
\begin{eqnarray}
P^{\mu} P_{\mu} | \Psi \rangle &=& M^2 | \Psi \rangle \ \label{eq:LFHamiltonianEquation} ,
\end{eqnarray}
where $ P^{\mu} $ is the energy-momentum four-vector operator. In the LF coordinates, the mass-squared operator, 
\begin{align}
H_{LC} \equiv P^2 = P^{\mu} P_{\mu}=P^+P^- - (P^{\perp} )^2 \ , \label{eq:MassSquared}
\end{align}
is analogous to the Hamiltonian in non-relativistic quantum mechanics. The details of the LF convention and notation in this work can be found in Refs. \cite{Li:2017mlw,Zhao:2013cma,Harindranth}. 
Since $P^+$ and $(P^{\perp} )^2$ are kinematical, the $P^-$, 
\begin{align}
P^- = & \frac{(P^{\perp} )^2 +M^2}{P^+} \ ,
\end{align}
is also referred to as LF Hamiltonian that generates the LF time-evolution (dynamics). In principle, $P^-$ can be obtained from a Lagrangian with a Legendre transformation.  

$H_{LC}$ can be numerically evaluated when expressed as a matrix eigenvalue problem in a complete set of basis as in BLFQ. In principle, the set of basis has infinite dimension. In practice, one limits the basis size by introducing truncation scheme(s). The resulting finite-dimensional eigenvalue problem can be evaluated numerically as a function of cutoff(s) in the truncation scheme(s). By extrapolation to the continuum limit, the physical observables can be obtained. 

\subsection{LF Hamiltonian density by Legendre transformation}
Treating the chiral Lagrangians via the LF formalism (see, e.g. Refs. \cite{Tsirova:2010zza,Mathiot:2013xta,Ji:2009jc,Ji:2013bca} and references therein) would usually result in the difficulty in solving the constraint equation of the nucleon field. In order to solve this difficulty, Miller \cite{Miller:1997cr,Miller:2000kv} suggested a chiral transformation of the field variables to obtain the chiral Lagrangian of the G\"ursey-type linear representation \cite{Chang:1967zza}. In this work, we follow Refs. \cite{Miller:1997cr,Miller:2000kv} and adopt a chiral model of the $N\pi$ system. 
The Lagrangian reads
\begin{align}
\mathcal{L} =  \frac{1}{4} f^2 \text{Tr}  \Big( \partial _{\mu} U \ \partial ^{\mu} U^{\dagger}  \Big)  & + \frac{1}{4} M^2_{\pi} f^2 \text{Tr} \Big( U + U^{\dagger} -2 \Big)  \nonumber \\
 & +  \bar{\chi}  \Big\{ \gamma _{\mu} i \partial ^{\mu} -M_N - M_N (U-1) \Big\} \chi    \ \label{eq:NewL} ,
\end{align}
which is a linear realization of the chiral symmetry. $f$ is chosen to be the pion decay constant (set as 93 MeV in this work). $M_N$ and $M_{\pi}$ are the nucleon mass and pion mass, respectively. $\chi$ denotes the bi-spinor field of the nucleon. $U$ is the unitary matrix for the chiral transformation, in which the pion field is introduced. 
If one works up to the order of ${1}/{f^2}$, $U$ takes the form \cite{Miller:1997cr,Miller:2000kv}
\begin{eqnarray}
U &=& 1 + i \gamma _5 \frac{\vec{\tau} \cdot \vec{\pi}}{f} - \frac{1}{2f^2} \pi ^2 + \mathcal{O}\Big(\frac{1}{f^3} \Big) \label{eq:U2} \ ,
\end{eqnarray}
where $\vec{\tau}$ denotes the Pauli matrices $\tau _a$ $(a=1,2,3)$, while $\vec{\pi}$ represents the scalar pion fields $\pi _{a}\ (a=1,2,3)$. 

The corresponding constraint equation of the nucleon field is 
\begin{align}
\chi _- &= \frac{1}{p ^+} \gamma ^0 \Big[  \gamma ^{\perp} \cdot p ^{\perp} + M_N  {U} \Big] \chi _+ \ ,
\end{align}
where the kinematic (dynamical) nucleon field component is $\chi _- $ ($\chi _+$).

By Legrendre transformation, we obtain the LF Hamiltonian density from Eq. \eqref{eq:NewL}. In this work, we keep only the terms that correspond to the processes of single-pion emission/absorption (up to the order of ${1}/{f}$). The resulting LF Hamiltonian density is 
\begin{align}
\mathcal{P}^- =  \frac{1}{2} \partial ^{\perp} \pi _a \cdot \partial ^{\perp} \pi _a & + \frac{1}{2} M_{\pi} ^2 \pi _a \pi _a  + \chi _+ ^{\dagger} \frac{(p^{\bot})^2 +M_N^2 }{p^+} \chi _+  \nonumber \\
& +  \chi _+ ^{\dagger} \Big[ - \gamma ^{\bot} \cdot i \partial ^{\bot} + M_N \Big] \frac{1}{p^+} M_N \Big[ i \gamma _5 \frac{\vec{\tau} \cdot \vec{\pi}}{f}  \Big] \chi _+ \nonumber \\ 
& + \chi _+ ^{\dagger} M_N \Big[  - i \gamma _5 \frac{\vec{\tau} \cdot \vec{\pi}}{f} \Big] \frac{1}{p^+} \Big[ \gamma ^{\perp} \cdot i \partial ^{\perp} + M_N \Big] \chi _+  + \mathcal{O}(1/f^2)  \ \label{eq:LFHorder1f}.
\end{align}
Higher-order contributions to $\mathcal{P}^-$ are expected to be corrections to the current calculation.


\subsection{Basis construction and truncation schemes}
\subsubsection{Symmetries}
The methodology of constructing the basis for carrying out the matrix eigenvalue solution of the LF mass-squared operator $H_{\rm LC}$ within a basis representation, BLFQ, is discussed in Refs. \cite{Vary:2009gt,Zhao:2013cma,Wiecki:2014ola}. In constructing the basis, we need to pay specific attention to the symmetries of the LF Hamiltonian $P^-$. These symmetries are: (1) the translational symmetry in the longitudinal direction, which results in the conservation of the total longitudinal momentum $P^+$; (2) the rotational symmetry in the transverse direction, which means that the longitudinal projection of the total angular momentum is conserved; (3) the conservation of net fermion number; and (4) transverse boost invariance. In this work, we also assume rotational symmetry in isospin space, where the longitudinal projection of the isospin of the constituent system is conserved. We construct the LF basis set according to these symmetries.

\subsubsection{Single particle basis}
We start with constructing the single-particle (s.p.) basis. In the longitudinal direction, we employ the discretized plane wave basis $\{ | p ^ + \rangle \}$. In particular, we constrain a particle in a longitudinal box of length $x_+=L$ and apply the periodic (anti-periodic) boundary condition to boson (fermion). The longitudinal momentum is discretized as
\begin{align}
p^+ = \frac{2 \pi}{L} j \ \label{eq:singleParticleRelation1}, 
\end{align}
where $j = 1, 2, 3, \cdots$ for bosons and $j= \frac{1}{2}, \frac{3}{2}, \frac{5}{2}, \cdots$ for fermions. Note that we exclude the ``zero modes" ($j=0$) for bosons (pions in this work). 

It is useful to define the longitudinal momentum fraction $x$ in terms of the total longitudinal momentum $P^+$ as
\begin{align}
x \equiv \frac{p^+}{P^+} \ = \ \frac{j}{K} \ \label{eq:singleParticleRelation2} ,
\end{align}
where the dimensionless parameter $K$ is related to $P^+$ via the relation $P^+ = \frac{2\pi}{L}K$. 

In the transverse direction, we employ the two dimensional harmonic oscillator (2DHO) basis. This choice of basis is useful to insure the transverse boost invariance of the LF kinematics \cite{Vary:2009gt,Brodsky:1997de}. The generating operator for the 2DHO basis can be expressed as \cite{Wiecki:2014ola}
\begin{align}
P_+ ^{\Omega} =& \frac{(p^{\perp})^2}{2p^+} + \frac{1}{2} {\Omega}^2 p^+ (r^{\perp})^2 \ = \ \frac{1}{2} \Omega \Big[ \frac{(p^{\perp})^2}{xP^+ \Omega} + x P^+ \Omega (r^{\perp})^2 \Big] \ ,
\end{align}
where the oscillator energy $\Omega$ is related to the energy scale of the 2DHO basis set as
\begin{align}
b= \sqrt{P^+ \Omega} \ .
\end{align}
In the following, we refer to $b$ as the basis strength. 

For the convenience in evaluating integrals involving 2DHO basis, we further introduce the momentum fraction weighted variables \cite{Maris:2013qma} as
\begin{align}
q^{\perp} \equiv  \frac{p^{\perp}}{\sqrt{x}}  , \ s^{\perp} \equiv \sqrt{x} r^{\perp} \ \label{eq:MomentumFractionWeightedVariables},
\end{align} 
where $[s_i^{\perp}, q_j^{\perp}]=i\delta _{ij}$ ($i,j=1,2$)  holds. The generating operator of the 2DHO basis in terms of the conjugate variables $(s^{\perp}, q^{\perp})$ can be rewritten as
\begin{align}
P_+ ^{\Omega} =& \frac{1}{2} \Omega \Big[ \Big(  \frac{q^{\perp}}{\sqrt{P^+ \Omega}} \Big)^2 + \Big( \sqrt{P^+ \Omega} s^{\perp} \Big)^2 \Big] \ .
\end{align}
In the momentum representation, the 2DHO wave function is
\begin{align}
\langle q^{\perp} | nm \rangle \ = \ \Psi _n^m (q^{\perp})\ = \ \frac{1}{b} \sqrt{\frac{4 \pi n!}{(n+|m|)!}} \rho ^{|m|} e^{-\frac{1}{2} \rho ^2} L_n ^{|m|} (\rho ^2)\ e ^{im \phi} \ ,
\end{align}
where the transverse momentum in the complex representation is
\begin{align}
& q^{\perp} = b \rho e^{i \phi} \ , \ (q^{\perp})^{\ast} = b \rho e^{-i \phi}
\end{align}
with $ \phi = \mathrm{arg}\ q^{\perp}, \ |q^{\perp}| = b \rho $. $n$, $m$ are the quantum numbers for the radial part and angular part of the wave function, respectively. They are related to the eigenenergy of the corresponding 2DHO wave function
\begin{align}
E_{nm} =& (2n+ |m|+1) \Omega \ .
\end{align}

In addition to the momentum space, we also have the the spin and isospin degrees of freedom for the $N\pi$ model. The s.p. basis can thus be classified according to the following set of quantum numbers
\begin{align}
| \alpha \rangle = & | x, n, m, s, t \rangle \ ,
\end{align}
where $s$ denotes the helicity. $t$ denotes the longitudinal projection of the isospin of the particle. It is understood that the nucleons are of spin $\frac{1}{2}$ and isospin $\frac{1}{2}$, while pions are of spin $0$ and isospin $1$. 

\subsubsection{Multi-particle basis}
The multi-particle basis is constructed as a direct product of the s.p. bases ($ \otimes | \alpha \rangle $). According to the symmetries of $P^-$ for the $N\pi$ system, we require the quantum numbers for all the constituent particles (labeled by $i$) in the retained multi-particle basis states to satisfy the following relations
\begin{align}
\sum _i p^+_i = P^+,  \ \sum _i m_i + \sum _i s_i = M_J, \ \sum _i t_i =T_z,\ \sum _i n^i = N_f \ \label{eq:SymmetryIdentitiesForBasis}.
\end{align}
The first identity requires all the basis states to have the same total longitudinal momentum. It is equivalent to 
\begin{align}
\sum _i j_i = K \ {\rm or} \  \sum _i x_i = 1 \ , 
\end{align}
according to Eqs. \eqref{eq:singleParticleRelation1} and \eqref{eq:singleParticleRelation2} for the fixed box-length $L$ and the total longitudinal momentum $P^+$. The second identity in Eq. \eqref{eq:SymmetryIdentitiesForBasis} states the conservation of the longitudinal projection of the total angular momentum $M_J$, which is produced by the helicity $s$ and the longitudinal projection of the orbital angular momentum $m$ of each constituent particle. (Note, however, the total angular momentum $J$ is not a good quantum number in the LF basis states.) The third identity in Eq. \eqref{eq:SymmetryIdentitiesForBasis} states that the longitudinal projection of the total isospin $T_z$  or, equivalently, total charge for the system is conserved.  The last identity in Eq. \eqref{eq:SymmetryIdentitiesForBasis} refers to the conservation of the net fermion number $N_f$, with $n^i=1$ for a nucleon and $n^i=0$ for each pion.

\subsubsection{Truncation scheme}
We apply three truncations in this work. First, the number of Fock sectors for the $N\pi$ system is truncated at the nucleon plus one-pion sector
\begin{eqnarray}
| N _{\rm phys} \rangle  &=& a |N \rangle + b | N \pi \rangle \ \label{eq:fockSectorTruncation},
\end{eqnarray}
with the amplitudes being $a= \langle N | N _{\rm phys} \rangle $ and $b = \langle N \pi | N _{\rm phys} \rangle $. It is also possible to include higher Fock sectors, e.g., $| N \pi \pi \rangle $. However, we will postpone this to future work. According to the Fock sector truncation Eq. \eqref{eq:fockSectorTruncation}, we have the net fermion number $N_f=1$ for all the basis states.

Second, we cut off the total longitudinal momentum for the many-body basis state 
\begin{align}
K = K_{\rm max}  ,
\end{align}
which makes the number of the longitudinal modes finite \cite{Hornbostel:1988fb}. The longitudinal continuum limit can be approached at the limit of $K _{\rm max} \rightarrow  \infty $ for a given box length $L$. 

Third, we truncate the number of the modes in the transverse direction for the many-body basis states by restricting the number of maximal excitation quanta, $N_{\rm max}$, as
\begin{align}
\sum _i (2n_i + |m_i| +1) \leq N_{\rm max} \label{eq:NmaxTruncation} ,
\end{align}
where $i$ denotes the constituent particles. By taking $N_{\rm max} \rightarrow  \infty $, the continuum limit in the transverse direction is realized.

\subsubsection{UV and IR cutoffs}
There are intrinsic ultravilot (UV) and infrared (IR) cutoffs imposed by the truncation in the transverse direction. For the 2DHO basis, the UV cutoff in momentum space is around $p^{\perp}_{\rm max} \propto b\sqrt{N_{\rm max}} $, while the IR cutoff is around $p^{\perp}_{\rm min} \propto b/\sqrt{N_{\rm max}}$.

\subsubsection{Factorization}
The application of the 2DHO s.p. basis in the transverse direction with $N_{\rm max}$ truncation admits an exact factorization of the LFWF into the ``intrinsic" and the ``center of mass" (CM) components \cite{Vary:2009gt,Caprio:2012rv, Barrett:2013nh}. Taking advantage of this factorization, the spurious CM excitation due to the adoption of the 2DHO s.p. basis can be eliminated by the use of a Lagrange multiplier term as explained below. The analogous factorization scheme has been adopted in the studies of nuclear structures (c.f., Ref. \cite{Caprio:2012rv,Barrett:2013nh}), where the three dimensional harmonic oscillator basis is adopted.

\subsection{Mode expansions}
The pion field can be expressed in terms of the creation and annihilation operators
\begin{align}
\pi _a( x) = & \sum _{k ^+} \sum _{\lambda = -1}^{\lambda =1}  \frac{1}{2\pi \sqrt{2Lk^+}} \int \frac{d^2 k^{\perp}}{(2\pi)^2} \Big[ a(k, \lambda) \varepsilon _a (\lambda) e^{-ikx} + a^{\dagger} (k,\lambda) {\varepsilon _a}^{\ast} (\lambda) e^{ikx} \Big] \ \label{eq:pionModeExpansion},
\end{align}
where we introduce the following polarization vectors to track the isospin degree of freedom of the scalar pion field $\pi _a \ (a=1,2,3)$
\begin{align}
\varepsilon (+1) = \frac{1}{\sqrt{2}}
\begin{pmatrix}
1 \\ 
i \\ 
0
\end{pmatrix} \ , \ 
\varepsilon (0) = 
\begin{pmatrix}
0 \\ 
0 \\ 
1
\end{pmatrix} \ , \ 
\varepsilon (-1) = \frac{1}{\sqrt{2}}
\begin{pmatrix}
1 \\ 
-i \\ 
0
\end{pmatrix} \ ,
\end{align}
with $\varepsilon ^{\dagger}(\lambda _i)\varepsilon (\lambda _j) = \delta _{\lambda _i, \lambda _j}$ and $\varepsilon (-\lambda) = {\varepsilon} ^{\ast} (\lambda)$. The subscript ``a" also indicates the component of the polarization vector $\varepsilon (\lambda)$. $\lambda$ denotes the longitudinal projection of the isospin of the physical pions, i.e., $\pi ^{\pm}$ and $\pi ^0$.

Similar to the pion field, the nucleon field can be represented with the creation and annihilation operators 
\begin{align}
\chi _+ (x) =& \sum _{p^+} \sum _{s ,t} \frac{1}{2\pi \sqrt{2L}} \zeta (s) T(t) \int \frac{d^2 p^{\perp}}{(2\pi)^2} \Bigg[ b(p,s,t) e^{-ipx} + d^{\dagger}(p,-s,-t) e^{ipx} \Bigg] \ \label{eq:nucleonModeExpansion},
\end{align}
where
\begin{align}
\zeta (+\frac{1}{2}) = (1,0,0,0)^T,\ \zeta (-\frac{1}{2}) = (0,1,0,0)^T \ , \\
T(+\frac{1}{2}) = (1,0)^T, \ T(-\frac{1}{2}) = (0,1)^T \ .
\end{align}

With the discretized longitudinal momentum [Eq. \eqref{eq:singleParticleRelation2}], the commutation and anticommutation relations are
\begin{align}
[a(k, \lambda), a^{\dagger}(k', \lambda ')] =& (2\pi)^2 \delta ^{(2)}(k_{\perp}-k'_{\perp}) \delta _{\lambda , \lambda '} \delta _{x,x'} \ , \\
\{ b(p,s,t), b^{\dagger}(p',s',t') \} =& (2\pi)^2 \delta ^{(2)}(p_{\perp}-p'_{\perp}) \delta _{s ,s '} \delta _{t,t'} \delta _{x,x'} \ , \\
\{ d(p,s,t), d^{\dagger}(p',s',t') \} =& (2\pi)^2 \delta ^{(2)}(p_{\perp}-p'_{\perp}) \delta _{s ,s '} \delta _{t,t'} \delta _{x,x'} \ .
\end{align}
Note with our Fock space expansion [Eq. \eqref{eq:fockSectorTruncation}], the independent field for the anti-nucleon is not included. The canonical anti/commutation relations of the field operators are
\begin{align}
[\pi _a (x), \pi _b (y)]_{x^+=y^+} =& - \frac{i}{4} \epsilon (x^- - y^-) \delta ^{(2)} (x^{\perp}-y^{\perp}) \delta _{ab} \ , \\
\{ \chi _+ (x) , \chi _+ ^{\dagger} (y) \}_{x^+=y^+} =& \frac{1}{2}\gamma ^0 \gamma ^+ \delta (x^- - y^-) \delta^{(2)} (x^{\perp} - y^{\perp}) \ .
\end{align}
$\epsilon (x) = \theta (x) - \theta (-x)$ is the antisymmetric step function, where the step function is
\begin{align}
\theta (x) = 0 \ \ {\rm for} \ x \leq 0 \ ; \ \theta (x) = 1 \ \ {\rm for} \ x>0 \ .
\end{align}
The relations $\frac{\partial \epsilon (x)}{\partial x} = 2 \delta (x)$ and $|x|=x \epsilon (x) $ hold. For the representation of the gamma matrices in this work, we follow the convention in Refs. \cite{Li:2017mlw,Zhao:2013cma,Harindranth}.

The creation and annihilation operators in terms of the 2DHO basis with the momentum fraction weighted variables are
\begin{align}
a(x, k^{\perp},\lambda) =& \frac{1}{\sqrt{x}} \sum _{n,m} \Psi _n^m (\frac{k^{\perp}}{\sqrt{x}} ) \alpha (x, n,m, \lambda) \ , \\ 
b(x, p^{\perp}, s ,t) = & \frac{1}{\sqrt{x}} \sum _{n,m} \Psi _{n}^m (\frac{q^{\perp}}{\sqrt{x}}) \beta (x, n,m,s ,t) \ ,
\end{align}
with the anti/commutation relations being
\begin{align}
[\alpha (x,n,m,\lambda), \alpha ^{\dagger}(x',n',m',\lambda)] = & \delta _{x,x'} \delta _{n,n'} \delta _{m,m'} \delta _{\lambda , \lambda '} \ , \\
\{ \beta (x,n,m,s ,t) , \beta ^{\dagger} (x',n',m',s ',t')  \} =&  \delta _{x,x'}  \delta _{n,n'} \delta _{m,m'} \delta _{s , s '} \delta _{t,t'} \ .
\end{align}

\subsection{Mass-squared operator}
The adoption of the 2DHO s.p. basis in the transverse direction results in the inclusion of the spurious CM excitation within the mass spectrum. In order to eliminate the CM excitation in the BLFQ approach, we introduce a Lipkin-Lawson Lagrange multiplier term \cite{Gloeckner:1974sst,Lipkin:1958zza} to the mass-squared operator $H_{\rm LC}$ [Eq. \eqref{eq:MassSquared}]. The modified mass-squared operator is  
\begin{align}
H =&  \ H_{\rm LC} + \Lambda (H_{\rm CM} - 2b^2 I) \  \label{eq:HwithCMregulated},
\end{align}
where $\Lambda >0 $ is the Lagrangian multiplier. The intrinsic motion in the solutions is not influenced by this Lawson term $(H_{\rm CM} - 2b^2 I)$ due to the factorization of the LFWF in the 2DHO basis with $N_{\rm max}$ truncation. Since the mass spectrum of the intrinsic motion is only determined by the intrinsic part of the LFWF, it is independent of $\Lambda$. The CM motion is governed by
\begin{align}
H_{\rm CM} = \big( P^{\perp} \big)^2 + b ^4 \big( R^{\perp} \big)^2 ,
\end{align}
where the CM momentum and coordinate in the transverse direction are respectively
\begin{align}
P^{\perp} = \sum _i p^{\perp} _i , \ R^{\perp} = \sum _i x_i r^{\perp} _i .
\end{align}
In terms of momentum fraction weighted variables [Eq. \eqref{eq:MomentumFractionWeightedVariables}], these CM variables are
\begin{align}
P^{\perp} = \sum _i \sqrt{x_i} q^{\perp} _i , \ R^{\perp} = \sum _i \sqrt{ x_i} s^{\perp} _i .
\end{align}
$H_{\rm CM}$ satisfies the eigenequation
\begin{align}
H_{\rm CM} |nm \rangle =  (2n +|m| +1) 2 b^2 |nm \rangle \label{eq:COMSpectrum} ,
\end{align}
where $|nm \rangle $ is the eigenvector that corresponds to the eigenvalue $\mathcal{E} _{nm} = (2n +|m| +1) 2 b^2$. Based on Eq. \eqref{eq:COMSpectrum}, it is easy to see that  the states with CM excitation (i.e., states with $n \neq 0$ and $m \neq 0$) are lifted in the spectrum; only the states with the lowest CM mode (i.e., states with $n=m=0$) remain unshifted \cite{Maris:2013qma}. In general, the spectrum of $H$ is a set of equally spaced approximate copies \footnote {The copies are not exact copies since the addition of available quanta to the CM motion means the loss of available quanta in the relative motion.} (named as subspectra), with the spacing characterized by $2 \Lambda b^2$ for every additional excitation quanta in the CM degree of freedom. In practice, we choose $\Lambda$ to be sufficiently large such that the subspectra with different CM modes are well separated. 

Making use of the LF Hamiltonian density $\mathcal{P} ^-$ [Eq. \eqref{eq:LFHorder1f}] and the mode expansions for pion and nucleon fields [Eqs. \eqref{eq:pionModeExpansion} and \eqref{eq:nucleonModeExpansion}], we calculate the LF Hamiltonian $P^-$ and hence the mass-squared operator [Eq. \eqref{eq:MassSquared}] as 
\begin{align}
H_{\rm LC} = & P^+ \underbrace{ \big( P^-_{\rm KE_N} + P^-_{\rm KE_{\pi}} + P^-_{\rm int} \big)}_{P^-} - \big(P^{\perp} \big)^2 \ ,
\end{align}
where $P^-_{\rm KE_N}$ and $P^-_{\rm KE_{\pi}}$ are the contributions from a free nucleon and a free pion, respectively. $P^-_{\rm int}$ is the interaction term that describes the contributions from the one-pion absorption and emission processes.

\subsection{Observables}
In terms of the LF basis set $\{ | \xi \rangle \}$ (with $ |\xi \rangle \equiv | x_N, n_N, m_N, s_N, t_N; x_{\pi} , n_{\pi} , m_{\pi}  , s_{\pi}=0 , t_{\pi} \equiv \lambda \rangle $), the matrix of the modified mass-squared operator for the $N\pi$ system (Eq. \eqref{eq:HwithCMregulated}) can be constructed. By solving the eigenequation (via numerical matrix diagonalization)
\begin{align}
H | \Psi _i \rangle =& M^2 _i | \Psi _i \rangle \ ,
\end{align}
we obtain the eigenmass $M_i$ and the corresponding eigenvector  
\begin{align}
| \Psi _i \rangle  \equiv &\sum _{\xi}   C_i ( \xi )\ | \xi \rangle \ ,
\end{align}
with $C _i ( \xi ) = \langle \xi | \Psi _i \rangle $ being the LF amplitude corresponding to the basis state $ | \xi \rangle $. The summation is taken over the LF basis set $\{ | \xi \rangle \}$. The LFWF is made up by the LF amplitudes $\{ \langle \xi | \Psi _i \rangle \}$.

We can apply the LFWF to compute observables for the hadronic structure, such as the PDF, the elastic electric and magnetic form factors, and the spin decomposition. As a illustration, we calculate the PDF in this work. The investigation of other observables will be presented in the future work.


\subsubsection{PDF}
The probability to find a constituent nucleon with the longitudinal momentum fraction $x_N$ in the current $N\pi$ model is 
\begin{align}
f(x_{N}) = {\sum} ' C^{\ast} ( \xi  )  C ( \xi )  \ \label{eq:PDFequation} ,  
\end{align} 
where it is understood that $x_{\pi} = 1 - x_{N}$ due to the conservation of the longitudinal momentum. The primed sum in Eq. \eqref{eq:PDFequation} denotes that the sum is over all the quantum numbers except $x_N$.

\section{Results and discussions}
\label{sec:ResultsDiscussions}
In this work, we adopt the Fock-sector-dependent renormalization (FSDR) \cite{Hiller:1998cv,Karmanov:2008br,Karmanov:2010ih,Karmanov:2012aj} scheme. We numerically diagonalize the matrix of the modified mass-squared operator $H$ [Eq. \eqref{eq:HwithCMregulated}], in which process the bare nucleon mass is tuned in the matrix elements within the single nucleon sector. This process is iterative and continues until the square-root of the eigenvalue of the ground state (identified as the physical proton) matches the mass of the physical proton (taken as 938 MeV in this work).

The mass counterterm is introduced only to the single-nucleon sector. In the FSDR scheme, we expect the mass counterterm to compensate for the mass correction due to the radiative processes: the quantum fluctuation from the single-nucleon sector to the $N\pi$ sector and back again. On the other hand, the nucleon mass in the $N\pi$ sector remains as the physical value. We fix the pion mass at 137 MeV in the FSDR procedure.

\subsection{Mass spectrum of the $N\pi$ system}
We first study the dependence of the mass spectrum of the $N\pi$ system on the model space, which is determined by the truncation parameters, $N_{\rm max}$ and $K_{\rm max}$, and the basis strength, $b$. For convenience, we set $K_{\rm max}$ to be $N_{\rm max} + 1/2$ throughout this work. 

\begin{figure}[h]
\centering
\includegraphics[width=13cm]{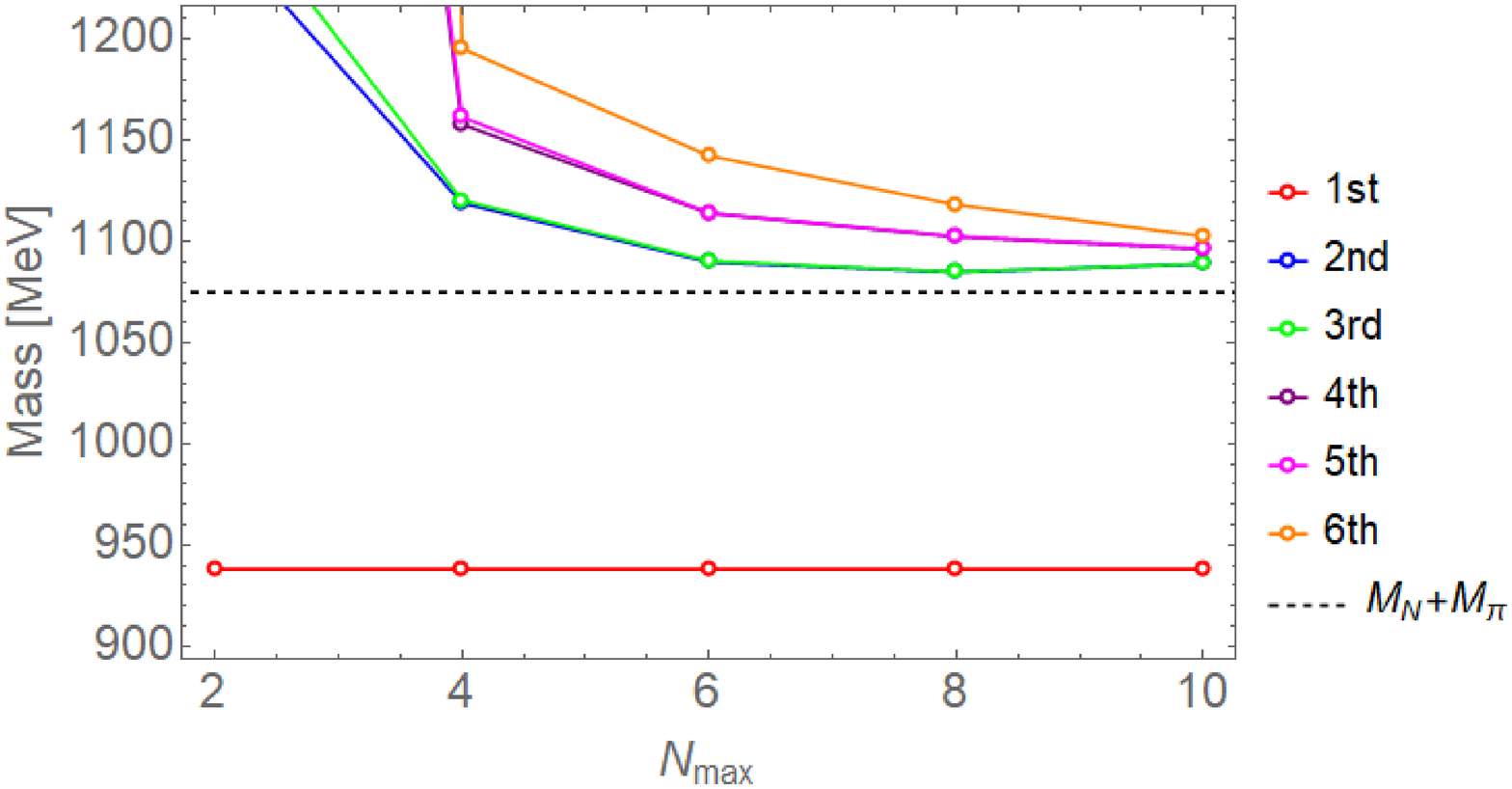}
\caption{Model space dependence of the spectrum of the $N\pi$ system computed via the BLFQ approach. The masses corresponding to the lowest 6 eigenstates are plotted as functions of $N_{\rm max}$ (set to be $K_{\rm max}-\frac{1}{2}$). The basis strength is fixed as $b=250$ MeV. The dashed line (at 1075 MeV) shows the threshold of the continuum of the $N\pi$ system. The ground (bound) state is identified as the physical proton.}
\label{fig:massSpectrum}
\end{figure}

In Fig. \ref{fig:massSpectrum}, we show the lowest 6 states in the mass spectrum of the $N\pi$ system as functions of $N_{\rm max}$, where we choose $b=250$ MeV as an example. We identify the ground (and also bound) state as the physical proton, of which the eigenvalue has been renormalized to 938 MeV by the FSDR procedure. The corresponding LFWF is boost invariant; it encodes all the information of the intrinsic structure of the proton. The other states appear to be the scattering states and their eigenvalues lie above the threshold of the continuum, which is the sum of the physical pion and proton masses adopted in this work (i.e., 1075 MeV).

We find all the eigenenergies of these 6 states seem to converge as $N_{\rm max}$ increases. The proof of the convergence is complicated and demanding in computing power; we will save the proof for the future work. As $N_{\rm max}$ increases, a better representation of the scattering states of the $N\pi$ system is anticipated. This can be inferred from the increasing level density of the scattering states as $N_{\rm max}$ increases.

\subsection{Proton's LFWF}
To compute the proton's LFWF, we need to fix the basis strength $b$ besides fixing the bare nucleon mass via the FSDR procedure for each choice of $N_{\rm max}$ ($K_{\rm max}=N_{\rm max} + 1/2$ as a reminder). This is achieved by varying $b$ to fit the r.m.s. charge radius of a proton $\sqrt{\langle r^2_{p,E} \rangle}$, which is 0.844 fm (see, e.g., Ref. \cite{Alarcon:2018zbz}). 
Overall, we fit for each $N_{\rm max}$  both the mass and the r.m.s. charge radius to respective physical values in order to determine the mass counterterm and $b$ in computing the proton's LFWF. In Table. \ref{table:parameterSettting}, we list the resulting model space parameters ($N_{\rm max}$ and $b$) to obtain the proton's LFWFs in this work. 

\begin{table}[h]
\caption{Model space parameters employed to obtain the proton's LFWFs. Note we set $N_{\rm max}=K_{\rm max}-\frac{1}{2}$.}
\centering
\begin{tabular}{c c c c} 
 \hline
 $N_{\rm max}$ & 6 & 8 & 10 \\ \hline
 $b$ [{\rm MeV}] & 176.95 & 245.54 & 279.55 \\ 
 \hline
\end{tabular}
\label{table:parameterSettting}
\end{table}

\subsection{Proton's PDF}

\begin{figure}[h]
\centering
\includegraphics[width=12cm]{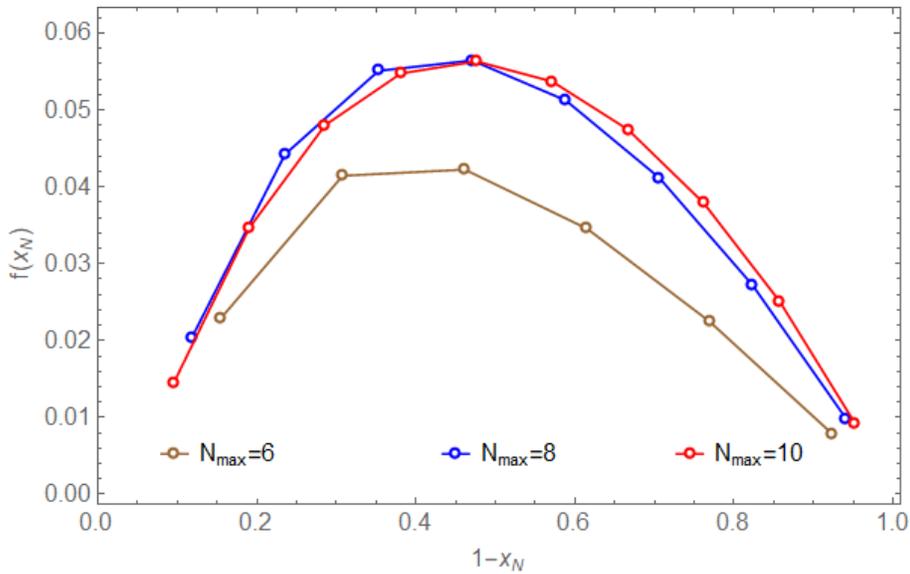}
\caption{The proton's PDF, $f(x_{N})$, as a function of the longitudinal momentum fraction of the constituent nucleon $x_N$ (note we rescale the $x$-axis as $1-x_{N}=x_{\pi}$ in the plot) and of the model space (defined by $N_{\rm max}$ and $b$). The details are in the text.
}
\label{fig:PDF_GS}
\end{figure}

We apply the proton's LFWF to compute its PDF, which encodes the distribution of the longitudinal momentum carried by its constituents. In this work, such PDF also represents the probability that a proton fluctuates into the constituent nucleon (of the longitudinal momentum fraction $x_N$) and pion (of the longitudinal momentum fraction $x_{\pi}$). 

In Fig. \ref{fig:PDF_GS}, the proton's PDF, $f(x_{N})$, is shown as a function of $x_{N}$ and the model space (with parameter settings shown in Table. \ref{table:parameterSettting}). Note we rescale the the $x$-axis as $x_{\pi}=1-x_N$ in the plot. We do not show the results for $f(x_{N}=1)$ in Fig. \ref{fig:PDF_GS}, which represents the probability to find a bare nucleon in the physical proton. For the cases with $N_{\rm max} =$ 6, 8 and 10, such probabilities are 0.83, 0.69, and 0.62, respectively.
  
For each $N_{\rm max}$, we verified that $f(x_{N})$ satisfies both the normalization condition and the momentum sum rule. As $N_{\rm max}$ increases, $f(x_{N})$ seems to converge (as indicated by the spacing between the curves and the positions of the peaks in the plot). Our results of $f(x_{N})$ peak at about $x_{\pi}=0.45$ (or $x_N=0.55$) for the model spaces with $N_{\rm max} =8$ and $10$. In the future, we plan to study $f(x_{N})$ where large/complete model spaces and high Fock sectors are applied. Also, the internal degrees of freedom of the constituents will be included to study the flavor asymmetry of the proton \cite{Aidala:2017ofy,Towell:2001nh,Hawker:1998ty,Alberg:2017ijg}.

\section{Conclusions and outlook}
\label{sec:ConclusionOutlook}
In this work, we apply, for the first time, the Basis Light-Front Quantization (BLFQ) method \cite{Vary:2009gt} to study a chiral model of the nucleon-pion ($N\pi$) system via an {\it ab initio}, non-perturbative, Hamiltonian approach. We demonstrate the approach with a test problem, in which the physical proton is treated as the relativistic bound state of the $N\pi$ system.

Starting from the Lagrangian density for the chiral model of the $N\pi$ system \cite{Miller:1997cr,Miller:2000kv}, we proceed with a Legendre transformation to obtain the corresponding light-front (LF) Hamiltonian density. In this work, we keep only the Fock sectors $|N \rangle$ and $|N\pi \rangle$. Correspondingly, we restrict the interaction terms in the LF Hamiltonian density and keep only the terms that correspond to the single-pion emission and absorption processes. 

We then show the construction and truncation schemes of our LF basis. As for the basis set in the momentum space, we employ the discretized plane wave basis in the longitudinal direction and the two dimensional harmonic oscillator basis in the transverse direction. Besides, we also discuss our basis construction in the spin and isospin degrees of freedom. We prune and truncate our basis according to the symmetry principles of our test problem.

We compute the matrix element of the mass-squared operator within our choice of the LF basis representation, where we also regulate the center of mass excitation by the Lipkin-Lawson method \cite{Lipkin:1958zza,Gloeckner:1974sst}. We obtain the mass spectrum of the $N\pi$ system and the corresponding boost-invariant light-front wave function (LFWF) by solving the eigenvalue problem of the resulting mass-squared matrix, in which process the mass counterterm is incorporated by the Fock-sector-dependent renormalization (FDSR) scheme \cite{Hiller:1998cv,Karmanov:2008br,Karmanov:2010ih,Karmanov:2012aj}. 

The mass spectrum of the $N\pi$ system in our solution includes both the bound and the  scattering states. We study the model space dependence of this spectrum. In particular, we investigate the eigenvalues of the lowest 6 states as a function of the model space, which is determined by the truncation parameters $N_{\rm max}$, $K_{\rm max}$, basis strength $b$, and the choice of Fock sectors. With increasing model space dimension, all the eigenvalues of these 6 states seem to converge, while the scattering states of the $N\pi$ system produce improving  representations of the continuum. Meanwhile, the eigenvalue of the ground state produces the physical proton mass for each model space with proper choice of the mass counterterm; such ground state is identified as the (physical) proton state. Note that larger Fock space would be necessary in order to verify the real convergence. We will postpone this verification to the future work.

To study the proton's parton distribution function (PDF), we compute the proton's LFWFs in a sequence of model spaces where both the proton's mass and its r.m.s. charge radius are fitted to respective physical values. For the resulting PDF, we investigate its dependencies on the model space and on the longitudinal momentum fraction of the constituent nucleon ($x_N$). We find the proton's PDF seems to converge with increasing model space dimension (scaled by $N_{\rm max}$). For the model spaces with $N_{\rm max}=8$ and $10$, the computed PDFs peak at about $x_N=0.55$ (or $x_{\pi}=0.45$). Further inclusion of the quark distribution functions of the constituent nucleon and pion could reveal the pion cloud's role in the light quark flavor asymmetry of the proton (see, e.g., Ref. \cite{Alberg:2017ijg}).

This work can progress into multiple paths in the future. We attempt to connect the current chiral model to the modern chiral effective theory (see, e.g., \cite{Entem:2003ft,Epelbaum:2008ga} and references therein). This work is currently ongoing. After this connection is accomplished, we plan to extend the current calculation (up to next-to-leading Fock sector) to incorporate systematically the contributions from higher Fock sectors, where we will examine the basis space dependence as well as the convergence of the Fock-sector expansion \cite{Li:2014kfa,Li:2015iaw}. We expect such investigations to be demanding in computing power. We plan to incorporate the technology of high performance computing (see Ref. \cite{Vary:2018hdv} and references therein). 

The current framework can be straightforwardly extended to investigate more nucleonic observables of great experimental interest, such as the transverse momentum distribution, and various categories of form factors. In addition, this framework can be extended to study more complicated nuclear systems, such as the deuteron, where the role of the relativistic dynamics is important but still unclear.

\section*{Acknowledgments}
We acknowledge valuable discussions with G. A. Miller, A. W. Thomas, E. Epelbaum, C. Weiss, M. Burkardt, P. Maris, T. Frederico, L. Geng, S. Jia, X. Ren and S. Tang. This work was supported by the U.S. Department of Energy (DOE) under grant No. DE-FG02-87ER40371. X. Zhao is supported by new faculty startup funding from the Institute of Modern Physics, Chinese Academy of Sciences.

\end{document}